\documentclass[aps,prb,twocolumn,superscriptaddress,floatfix]{revtex4-2}
\usepackage{graphicx}
\usepackage{amssymb}
\usepackage{amsfonts}
\usepackage{amsmath}
\usepackage{textcomp}
\usepackage{gensymb}

\usepackage{verbatim}
\usepackage{xcolor}
\usepackage{siunitx}
\usepackage{tikz}
\usepackage{subcaption}

\newcommand{\T}{\S\kern .15em\relax }

\newcommand{\um}{\ \text{$\mu$m}\ }

\begin{document}
\title{Visualization by scanning SQUID microscopy of the intermediate state in the superconducting Dirac semimetal PdTe${}_2$ }

\author{P. Garcia-Campos} \email{pablo.garcia-campos@neel.cnrs.fr} \affiliation{Universit\'{e} Grenoble Alpes \& Institut N\'{e}el, CNRS, 38042 Grenoble, France}

\author{Y. K. Huang} \affiliation{Van der Waals-Zeeman Institute, University of Amsterdam, Science Park 904, 1098 XH Amsterdam, Netherlands}
\author{A. de Visser} \affiliation{Van der Waals-Zeeman Institute, University of Amsterdam, Science Park 904, 1098 XH Amsterdam, Netherlands}
\author{K. Hasselbach} \email{klaus.hasselbach@neel.cnrs.fr} \affiliation{Universit\'{e} Grenoble Alpes \& Institut N\'{e}el, CNRS, 38042 Grenoble, France}

\date{\today}

\begin{abstract}
The Dirac semimetal PdTe${}_2$ becomes superconducting at a temperature $T_{c}=1.6$~K. Thermodynamic and muon spin rotation experiments support type-I superconductivity, which is unusual for a binary compound. A key property of a type-I superconductor is the intermediate state which presents a coexistence of superconducting and normal domains at magnetic fields lower than the thermodynamic critical field $H_{c}$. We present Scanning SQUID microscopy (SSM) studies of PdTe${}_2$ revealing coexisting superconducting  and normal domains of tubular and laminar shape as the magnetic field is more and more increased thus confirming type-I superconductivity in PdTe${}_2$. Values for the domain wall width in the intermediate state have been derived. The field amplitudes measured at the surface indicate bending of the domain walls separating normal and the superconducting domains.
\label{key}

\end{abstract}

\maketitle

\section{INTRODUCTION}

Finding materials presenting topological superconductivity is an important challenge in today's condensed matter research. Topological superconductors are predicted to host Majorana zero modes at their surface, which could be used for quantum computation with increased coherence times because the surface states are protected by symmetry~\cite{Qi-Zhang2010,sarma:majorana:2015}. A wide range of unconventional superconductors are under scrutiny for signs of topologically protected states~\cite{Ando-Fu2015,Sato-Fujimori2016,Sato-Ando2017}. A promising family of materials are the transition metal dichalcogenides, to which PdTe${}_2$ belongs. Angle resolved photoemission spectroscopy (ARPES) has identified PdTe${}_2$ as a Dirac semimetal, with a tilted Dirac cone below the Fermi energy with spin-polarized topological surface states~\cite{Liu:2015,Clark,Bahramy2018}. Since the tilt parameter $k > 1$, PdTe${}_2$ is classified as a type-II Dirac semimetal~\cite{Soluyanov2015}. PdTe${}_2$ is also a superconductor below $T_c = 1.6$~K~\cite{Guggenheim1961}, with a conventional fully-gapped order parameter indicated by the step in the specific heat at $T_c$, $\Delta C /\gamma T_{c}\approx 1.5~$~\cite{Amit} ($\gamma$ is the Sommerfeld coefficient), and supported by the exponential temperature variation of the London penetration depth~\cite{Salis,Teknowijoyo2018}. Scanning tunneling microscopy/spectroscopy (STM/STS)~\cite{Das,Sirohi:2019,Clark} and point contact spectroscopy (PCS)~\cite{Le} measurements report a BCS gap size $\Delta_{BCS}$ of the order of $215-326~\mu$eV, which gives rise to $2 \Delta_{BCS}/$k$_BT_c$ in the range $3.0-4.2$,\textit{ i.e.} close to the weak coupling value of 3.52.

The superconducting state of PdTe${}_2$ in applied magnetic fields is subject of debate. Dc-magnetization and ac-susceptibility measurements show the presence of the differential paramagnetic effect (DPE) in applied magnetic fields $(1-N)H_c < H_a < H_c$, where $\mu_0 H_c = 13.6$~mT is the thermodynamic critical field~\cite{Carley} and $N$  the demagnetization factor of the single crystal used in the experiment. This provides strong evidence for the existence of the intermediate state, which is characterized by a macroscopic phase separation in superconducting and normal domains and which is a key property of a type-I superconductor~\cite{Tinkham}. Type-I superconductivity is in-line with the reported value of the Ginzburg-Landau parameter $\kappa=\lambda/\xi \approx 0.09-0.34$~\cite{Salis,Carley}, where $\lambda$ is the magnetic penetration depth and $\xi$ the superconducting coherence length. This value of $\kappa$ is smaller than the theoretical boundary value $1/\sqrt{2}$, above which type-II behavior is expected. On the other hand, STM/STS~\cite{Das,Sirohi:2019} and PCS~\cite{Le} experiments have given rise to an interpretation in terms of a mixed type-I and type-II superconducting phase along with a spatial distribution of critical fields. This was attributed to an intrinsic electronic in-homogeneity already present in the normal phase. In another STM/STS measurement~\cite{Clark} the observation of a vortex core and type-II superconductivity is reported. However, in all these STM/STS experiments an Abrikosov vortex lattice, which is the hall mark of type-II superconductivity~\cite{Tinkham}, was not observed. More recently, transverse muon spin rotation ($\mu$SR) measurement have been conducted to probe the intermediate state on the microscopic scale~\cite{Leng2019}. The results provide solid evidence for type-I superconductivity in the bulk of the PdTe${}_2$ crystal.

These conflicting results and their interpretation provide the motivation to study the magnetic flux structure in the superconducting phase at the local scale. Here we report local magnetization measurements in the superconducting phase of a PdTe${}_2$ single crystal using a scanning SQUID microscope~\cite{hykel:microsquid:2014}. Thus the focus of the present paper is on the nature of the superconducting state (Type-I or Type-II), rather than on aspects of topological superconductivity.

\section{EXPERIMENTAL METHODS}

Our measurements were made with a high-resolution scanning $\mu$-SQUID microscope (SSM) working in a dilution refrigerator~\cite{Veauvy:RSI:2002,hykel:microsquid:2014}.

 The critical current, $I_{c}(B)$, of the $\mu$-SQUID is a periodic function of the flux, $\Phi$, penetrating the SQUID loop, with a period equal to the magnetic flux quantum, $\Phi_0=h/2e$. By measuring the critical current 600 times per second we achieve a flux resolution of $1.2 \times 10^{-4}~ \Phi_0/\sqrt{\textrm{Hz}}$.
The square shaped aluminum $\mu$-SQUID has an effective area of $S_{SQUID} = 0.36~\mu$m$^{2}$, thus a magnetic induction B of 5.7~mT threads 1~$\Phi_0$ of flux through the $\mu$-SQUID, see Fig.~\ref{fig:PdTe2:phasediagram}(a).

SQUID microscopy and tuning fork based force microscopy are combined in this microscope. The $\mu$-SQUID is situated at the very tip of a silicon chip. Mounting the $\mu$-SQUID chip on a piezoelectric quartz tuning fork allows to maintain contact between the chip and the sample surface while scanning ~\cite{Veauvy:RSI:2002}. The SQUID-sample height is obtained by measuring the distance of the SQUID on the silicon chip relative to the tip's apex and the angle between the SQUID chip and the sample using a microscope equipped with a camera. For an angle of 4$\degree$ and 2 $\mu$m SQUID-tip distance a SQUID-sample height of 150~nm is obtained. Measurements are made at a safe height of an additional 200~nm above the surface.

The microscope maps the SQUID's critical current as a function of the SQUID's position. For further data treatment the critical current maps are transformed to magnetic field maps using $I_{c}(B)$ calibration curves similar to the black trace of Fig.~\ref{fig:PdTe2:phasediagram}(a). Thus the images shown represent charts of the magnetic field above the sample surface.

The measurements were performed on a PdTe${}_2$ single crystal in the shape of a flat rectangular prism with a length 0.88~mm, width 0.84~mm and thickness 0.097~mm. 
With two others crystals this one was used prior for measurements of the London penetration depth, $\lambda(T)$, labeled s1 in Ref.~\onlinecite{Salis}. For all of the three samples the onset superconducting transition temperature was found to be 1.66 $\pm$ 0.02~K and the zero temperature penetration depth $\lambda (0) = 470 \pm 10 $~nm for $H$ $\perp c$-axis.
For the SSM measurements the applied field is directed along the crystal's $c$-axis. A demagnetization factor $N=0.788$ is calculated \cite{Aharoni1998}.

\section{Results}

\begin{figure*}[h!tb] 
\centering
\includegraphics[scale=0.98]{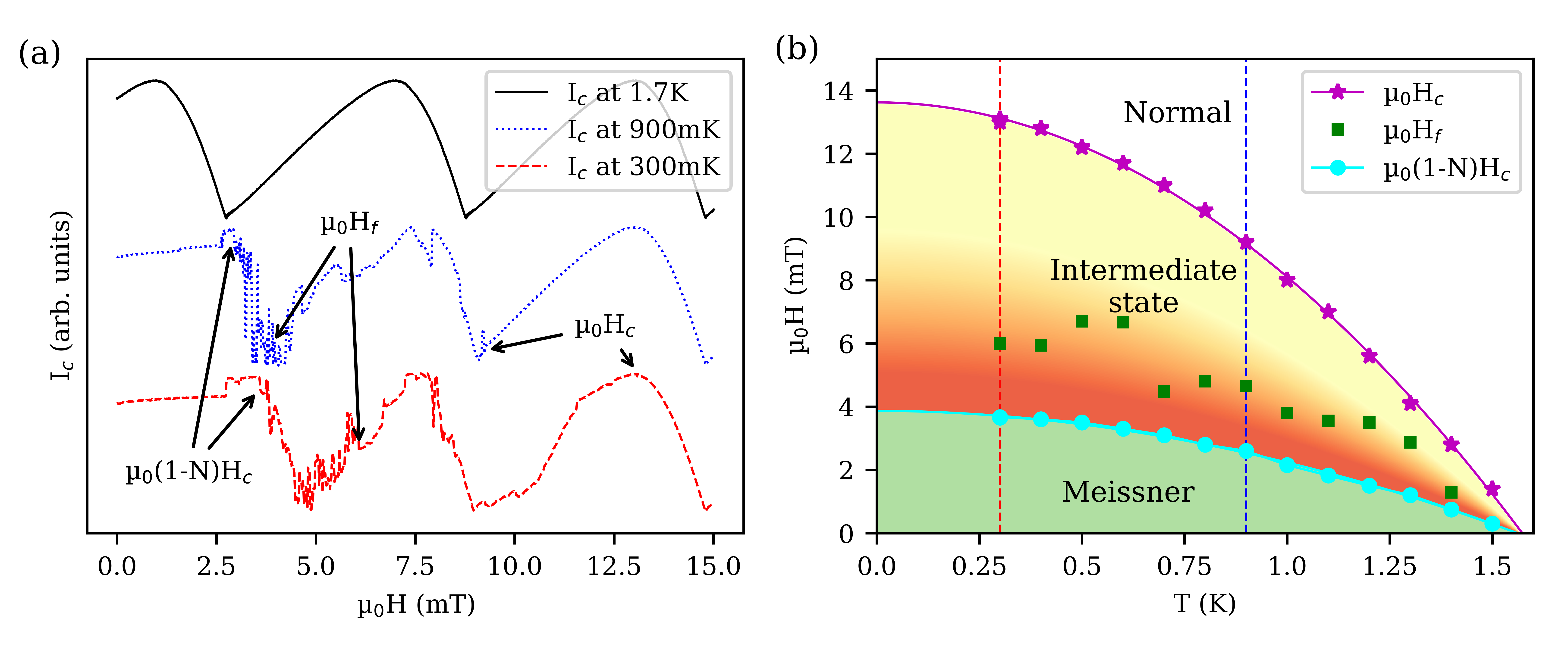}
\caption{(Color online) (a) The SQUID response after ZFC at three temperatures as indicated. In the normal phase (black solid line) the SQUID response is smooth. In the superconducting phase we distinguish three different behaviors: (\textit{i}) flat response, \textit{i.e.} screening for $H<(1-N)H_c$, (\textit{ii}) high density of I$_{c}$ jumps, \textit{i.e.} penetration of magnetic flux tubes for $(1-N)H_c<H<H_f$, and (\textit{iii}) smoother jumps, \textit{i.e.} when flux tubes fuse into laminar structures $H_f<H<H_c$. The fields $H_p=(1-N)H_c$, $H_f$ and $H_c$ are indicated by arrows. In (b) the phase diagram is constructed from the gathered characteristic field values. The solid magenta line represents a BCS-fit $H_c(T)=H_c(0)[1-(T/T_c)^2]$ with $\mu_0H_c(0)=13.62$~mT and $T_c=1.57$~K and the solid cyan line the equivalent fit with $\mu_0H_p(0)=3.83$~mT and $T_c=1.58$~K. The green squares indicate the field values when the flux changes become smoother above which laminar structures appear. The vertical dashed lines indicate the temperatures at which the SQUID response is shown in (a).}

\label{fig:PdTe2:phasediagram}
\end{figure*}

In order to investigate the $H-T$ phase diagram we have placed the SQUID at about 350 nm above the center of the sample. After zero field cooling (ZFC) we recorded the SQUID response on increasing the applied magnetic field, $H_a$, for a number of fixed temperatures. In Fig.~\ref{fig:PdTe2:phasediagram}(a), we show the critical current, $I_c$, as a function of $H_a$ for three temperatures: $T=1.7$~K (black line), 0.9~K (blue line) and 0.3~K (red line). At $T=1.7$~K the sample is in the normal state and the data shows the modulation (arcs) of the SQUID's critical current. Each period corresponds to one flux quantum entering the SQUID loop. At 0.9~K and 0.3~K the sample is in the superconducting state. The data starts off with a flat response, which corresponds to Meissner screening, up to a penetration field $H_p= (1-N)H_c$. Above $H_p$ the sample is in the intermediate state and flux penetrates in a rather abrupt manner, as indicated by the fluctuating signal. Above $H_c$ non-affected arcs are recorded, the sample is in the normal phase. 
The field values $(1-N)H_c$ and $H_c$ measured in this way are indicated by arrows in Fig.~\ref{fig:PdTe2:phasediagram}(a). In between these fields we denote a significant change in the SQUID response, from large to small fluctuations of $I_c$, at a \textit{fusing} field, $H_f$. As we will show in the next section, at this field tubular magnetic structures start fusing into laminar structures.

In Fig.~\ref{fig:PdTe2:phasediagram}(b) we have collected values of $(1-N)H_c$, $H_f$ and $H_c$ obtained at thirteen different temperatures. $H_c$ follows the standard BCS behavior, $H_c(T) = H_c(0)[1-(T/T_c)^2]$, with $\mu_0H_c = 13.62\pm 0.05$~mT and $T_c = 1.57\pm0.01$~K. These value are in excellent agreement with the $H_c(T)$ behavior reported in Ref.~\onlinecite{Carley}. Correspondingly, we obtain $\mu_0(1-N)H_c(0) = 3.83\pm0.03$~mT which gives us a demagnetization factor $N=0.72$. This value is smaller than the calculated one, $N=0.788$, which we attribute to a measured effective value $N_{eff} < N$ due to the local probe geometry. The effectiveness of this method for determining the phase diagram is the result of a very low resistance to flux penetration and weak flux pinning in PdTe${}_2$.

\subsection{PdTe$_2$ Zero Field Cooled}

\begin{figure*}[htb]
\centering
\includegraphics[width= 164mm]{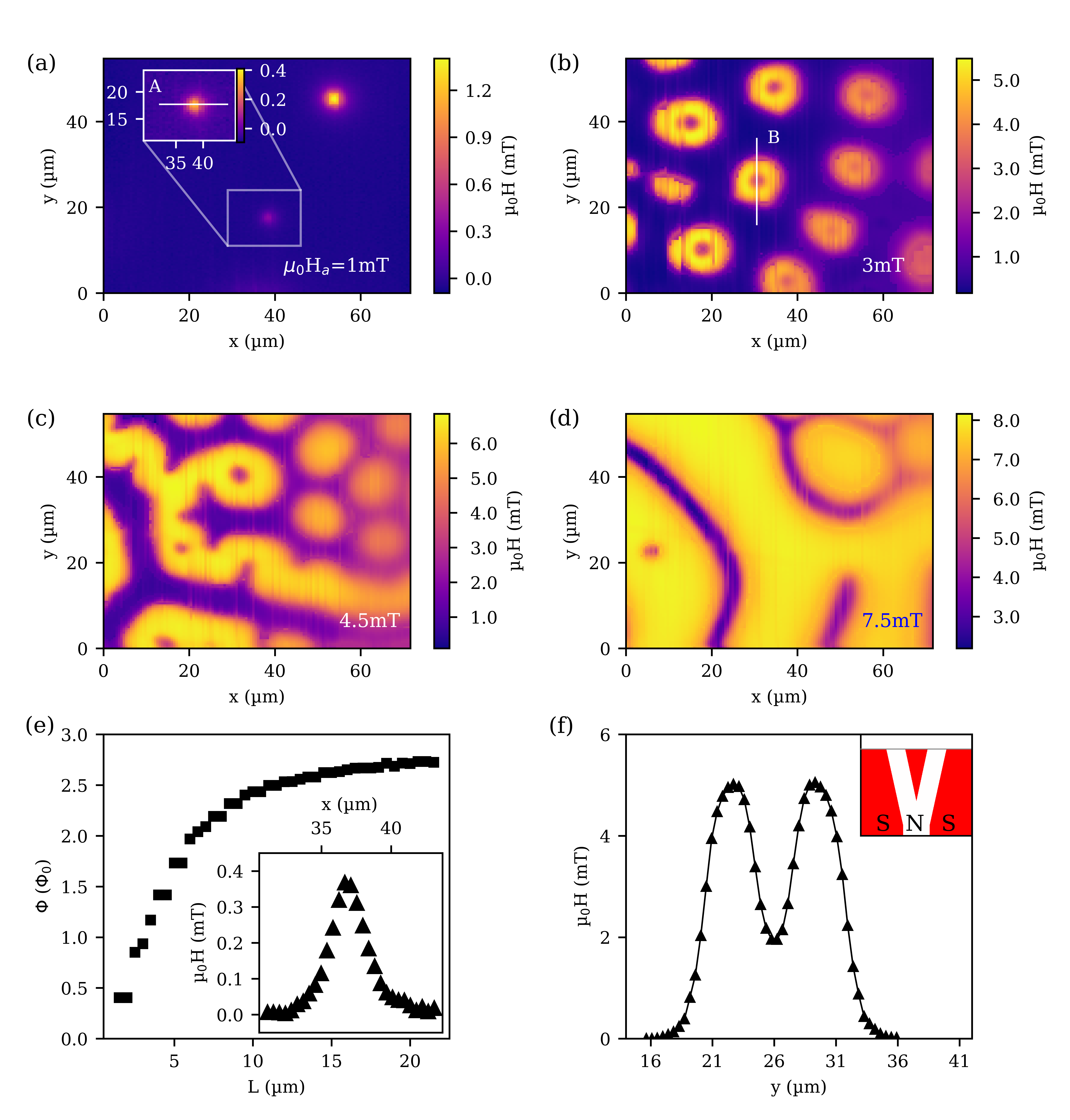}
\caption{(Color online) Panels (a) to (d): ZFC scanning SQUID images taken at a temperature of $900$~mK ($H_c=9.1$~mT) at an applied field of $1$~mT (reduced field, $h=H_a/H_c$, of 0.11), $3$~mT (0.33), $4.5$~mT (0.5) and $7.5$~mT (0.82), respectively (all images from the same cool down). These images show the magnetic flux structures in the different regions of the phase diagram (Meissner, intermediate tubular and intermediate laminar in Fig. \ref{fig:PdTe2:phasediagram}(b)), dark grey (blue) regions are superconducting and light grey (orange) normal. The panel (a) contains a zoom on the weakest flux tube we observed. In panel (e) the inset shows the flux profile along the line A of the flux tube in the zoom in panel(a), the main panel shows the increase in collected flux as the magnetic field is summed up over areas with increasing lateral length, L. and (f): Field profile along the line B as shown in (b). 
The inset in (f) represents the schematics of flux tube branching at the surface of the sample, neglecting NS interface bending.}
\label{fig:PdTe2:ZFCimages}
\end{figure*}

In order to investigate how the flux penetration develops in the intermediate state, we took magnetic images of the crystal at $T=0.9$~K (blue dashed line in Fig. \ref{fig:PdTe2:phasediagram}(b)) for different applied fields after ZFC (see Fig.~\ref{fig:PdTe2:ZFCimages}). At the lowest applied field $\mu_0 H_a = 1$~mT we expect flux exclusion, which is confirmed by the data in Fig.~\ref{fig:PdTe2:ZFCimages}(a). Nonetheless, some magnetic structures are observed, but since they do not evolve with the applied magnetic field, we conclude that they were created by the residual magnetic field upon cooling. The field profile of the smallest structure along line A in the zoom of Fig.~\ref{fig:PdTe2:ZFCimages} (a) is plotted in the inset of Fig.~\ref{fig:PdTe2:ZFCimages}(e).
We note that this structure is the least intense we found.

When $H_p$ is crossed, magnetic structures fill the space, as demonstrated by the images acquired at 3.5, 4.5 and 7.5~mT (see Fig.~\ref{fig:PdTe2:ZFCimages} (b,c,d)). We observe a weaker magnetic contrast on the right side of the sample, that we attribute to an increased tip sample distance due to a spurious contact between the SQUID sensor and a high point on the sample. As shape and density of the magnetic features are consistent over the images the processes driving the formation of the flux structures are not affected by this increased tip sample distance.

At 3.5~mT,  Fig.~\ref{fig:PdTe2:ZFCimages}(b), the intermediate state is established and a self-organized lattice of flux tubes is observed.
 We notice two types of magnetic structures with closed topology: \textit{mountains}, see the structures in Fig.~\ref{fig:PdTe2:ZFCimages}(a) and the profile in Fig.~\ref{fig:PdTe2:ZFCimages}(e), and \textit{volcanoes}, see Fig.~\ref{fig:PdTe2:ZFCimages}(b) and the profile in Fig.~\ref{fig:PdTe2:ZFCimages}(f).

A priori, these closed structures should obey flux quantization~\cite{Tinkham}, and the appearance of isolated single-$\Phi_0$ structures, such as reported in Refs.~\cite{Clark,Sirohi:2019}, cannot be excluded.
We quantified the amount of flux contained in the weakest flux structure by two methods. One is fitting the flux profile of a hypothetical Abrikosov vortex to the measured flux profile. As penetration depth and height above the surface are interdependent parameters \cite{Kogan2003a} of the stray field originating from an Abrikosov vortex, we fixed the SQUID height to 350~nm and obtained a good agreement for an effective penetration depth of $\lambda_{eff}$ = 2.1 $\mu$m and a total flux of $7\Phi_0$. The large $\lambda_{eff}$ compared to the $\lambda$ obtained
by Salis et al.~\cite{Salis} ($\lambda(0) H_{ac} \parallel c$-axis $\sim377$~nm) could be indicative for field spreading effect more important than in type-II superconductors.
The other method is model free, based only on the fact that the magnetic flux through an area is equal the integral of B over this area. Fig.~\ref{fig:PdTe2:ZFCimages}(e) shows the increase in collected flux as the area of integration (square with length of side L, centered at the flux spot in the zoom of Fig.~\ref{fig:PdTe2:ZFCimages}(a)) is increased. Before integration a linear plane fit was used to subtract any field offset. This methods tends to indicate $2.8\Phi_0$ for the amount of flux contained in the flux tube. As magnetic flux in a superconductor is quantized this indicates that the flux contained in the structure is $3\Phi_0$ and our procedure misses 10\% of the total flux.

The strong dip in the center, \textit{volcano}, of the profile in Fig.~\ref{fig:PdTe2:ZFCimages}(f) indicates the presence of a superconducting region as sketched in the inset. 
Landau predicted in 1938 that normal domains could carry superconducting inclusions in order to minimize electrostatic energy~\cite{Landau1938,Landau1943,Huebener,Tinkham}. This phenomenon is called branching and has been visualized ~\cite{Allen1974} using Bitter decoration. The residual magnetic field above the superconducting region in the center of the \textit{volcano} is attributed to the overlapping stray fields of the surrounding normal region.

The observation of regular tubular flux structures has been reported in the literature \cite{Huebener,Prozorov2005}, only for the cleanest type-I superconductors. Goren \textit{et al.} \cite{Goren1971} and Clem \textit{et al.} \cite{Clem2013} 
could predict the transition from tubular to laminar shapes of the flux structures evolving in size and density as a function of applied magnetic field for a given thermodynamic critical field. Clem \textit{et al.}\cite{Clem2013} proposed a coherent description from the low field region, describing normal tubes in the superconducting state, followed by intermediate fields, with laminar structures of alternating normal and superconducting regions, up to superconducting tubes surrounded by normal state regions close to the critical field. The optimal flux configuration is obtained in minimizing the sum of the excess energy of the nonuniform magnetic field in vacuum close to the sample and the positive wall energy due to the surfaces (S) between the superconducting and normal regions. The wall energy, $E_{w}$, is expressed as the superconducting condensation energy density multiplied by the surface area, $S$, times a domain wall width $\delta$, $E_{w}$= $\delta S B_{c}^{2}/ 2\mu_{0}$.

We first derive the domain wall width following Ref.~\cite{Goren1971}.
Being a function of reduced field, $h=H_a/H_c$ and sample thickness, $d$, the domain wall width can be obtained either from the flux spot diameter, $D$, via the relation $\delta=D{^2}(1-h)(1-\sqrt{h})/(2d)$, or from the lattice parameter between adjacent flux spots, $a$, via the relation $\delta=a{^2}h(1-h)(1-\sqrt{h})/(2d)$.
Choosing Fig.~\ref{fig:PdTe2:ZFCimages}(b) to measure the lattice parameter and sizes of spots ($D=13\pm2~\mu$m and $a=20.1\pm2.5~\mu$m), we obtain a domain wall width of $0.24\pm0.08~\mu$m based on the spot size and $0.19\pm0.06~\mu$m based on the lattice parameter, respectively for $h=3/9.1$ and $d=97~\mu$m. A second way to calculate $\delta$ is given by Clem \textit{et al.}, who add in their expression for $\delta$ a normalized free energy, $\Phi_{1}$, that attains the value $\Phi_{1} =0.079$ for this field. Based on the spot size diameter, the domain wall width can be calculated from the relation
$\delta={(D\Phi_{1}/h)}^{2}/ d$ and we obtain $\delta = 0.10\pm0.06~\mu$m.
Alternatively, Clem \textit{et al.} estimate $\delta$ from a normalized lattice parameter $R_{0} = {(\sqrt{3}/2\pi)}^{1/2}a = 11.9~\mu$m with help of the relation
$\delta={(2R_{0}\Phi_{1})}^{2}/(dh)$. This results in $\delta=0.11\pm0.03~\mu$m. Since the model of Clem \textit{et al.} does neither take into account the spreading of the flux tubes near the surface nor the branching of the flux tubes that we observe, we argue the most reliable estimates of the domain wall width are the ones based on the lattice parameter, thus $\delta = 0.11 \pm0.03~\mu$m according to the model of Clem \textit{et al.} or $0.19\pm0.08~\mu$m according to the model of Goren \textit{et al.}

Above a certain threshold field, $H_f$, the tubular magnetic structures fuse into laminar domains, as for instance is shown by the scan taken at 4.5~mT reported in Fig.~\ref{fig:PdTe2:ZFCimages}(c). The values $H_f(T)$ (green squares) in Fig.~\ref{fig:PdTe2:phasediagram}(b) show a relatively high dispersion which we attribute to the coexistence and competing effects of structures with a closed (tubular) and an open (laminar) topology. Such coexistence of shapes has been reported in the literature before, and a quantitative analysis has been made in several model cases~\cite{Huebener,Brandt2011,Berdiyorov,Goren1971} and is consistent with the small free energy difference between the flux arrangements~\cite{Clem2013}.

The domain wall width is in all configurations an important parameter. Using Ref.~\cite{Clem2013} we have derived also the domain wall width in the laminar state observed at 4.5~mT (Fig.~\ref{fig:PdTe2:ZFCimages}(c)). The distance between two normal laminae is $2R_{0}=20~\mu$m and the width of the laminae is $2R=8~\mu$m. The normalized free energy at the reduced field $h = 4.5/9.1$ is $\Phi_{2}$ = 0.092.
The domain wall width infered from the period of the normal laminae is expressed as
$\delta$=${(4R_{0}\Phi_{2})}^{2}/d$, and we obtain $\delta = 0.14~\mu$m. On the other hand, the domain wall width derived from the width of the normal laminae is expressed as
$\delta={(4R\Phi_{2}/h)}^{2}/d$, which leads to a value $\delta= 0.09~\mu$m.

It is remarkable that the model of Clem \textit{et al.} gives consistent results of the order of 0.1~$\mu$m for $\delta$ for as well the tubular as the laminar state, considering the level of abstractness of the model compared to the complex shapes observed in real samples.

As the field increases, the normal laminae become wider and occasionally some tubular regions are observed, for instance at ($x=5~\mu$m, $y=22~\mu$m) in the scans of Fig.~\ref{fig:PdTe2:ZFCimages}(d). This tells us that the high field equilibrium state in our case is a mixture of tubular and laminar superconducting structures at odds with exclusively tubular structures predicted for the high field phase in the Clem \textit{et al.} model.

At low and intermediate fields we observe only one single funnel-like branching per normal domain. Similar branching pattern to those in Fig.~\ref{fig:PdTe2:ZFCimages}(c) have been reported~\cite{Huebener,Prozorov2005} in the case of elemental type-I superconductors.

Finally, we remark that branching is expected to occur only for a sample thickness larger than the critical thickness $d_s \approx 800 \times \delta $~\cite{Huebener}. With our estimate of the domain wall width in the range 0.1 to $0.2~\mu$m, $d_s$ falls in the range $80-200~\mu$m, while the sample thickness is $97~\mu$m. Thus observation of branching is in favor of a domain wall width of the order of 0.12~$\mu$m.

The partial duplication of the structure at x=15\um in Fig.~\ref{fig:PdTe2:ZFCimages}(b) and the vertical lines in Fig.~\ref{fig:PdTe2:ZFCimages}(b)-(d) denote movement of the structures, which we attribute to the coupling between the SQUID's magnetic field and the structure itself. This movement can only be observed in case of weak pinning.

\begin{figure}[htb]
\centering
\includegraphics[scale=0.95]{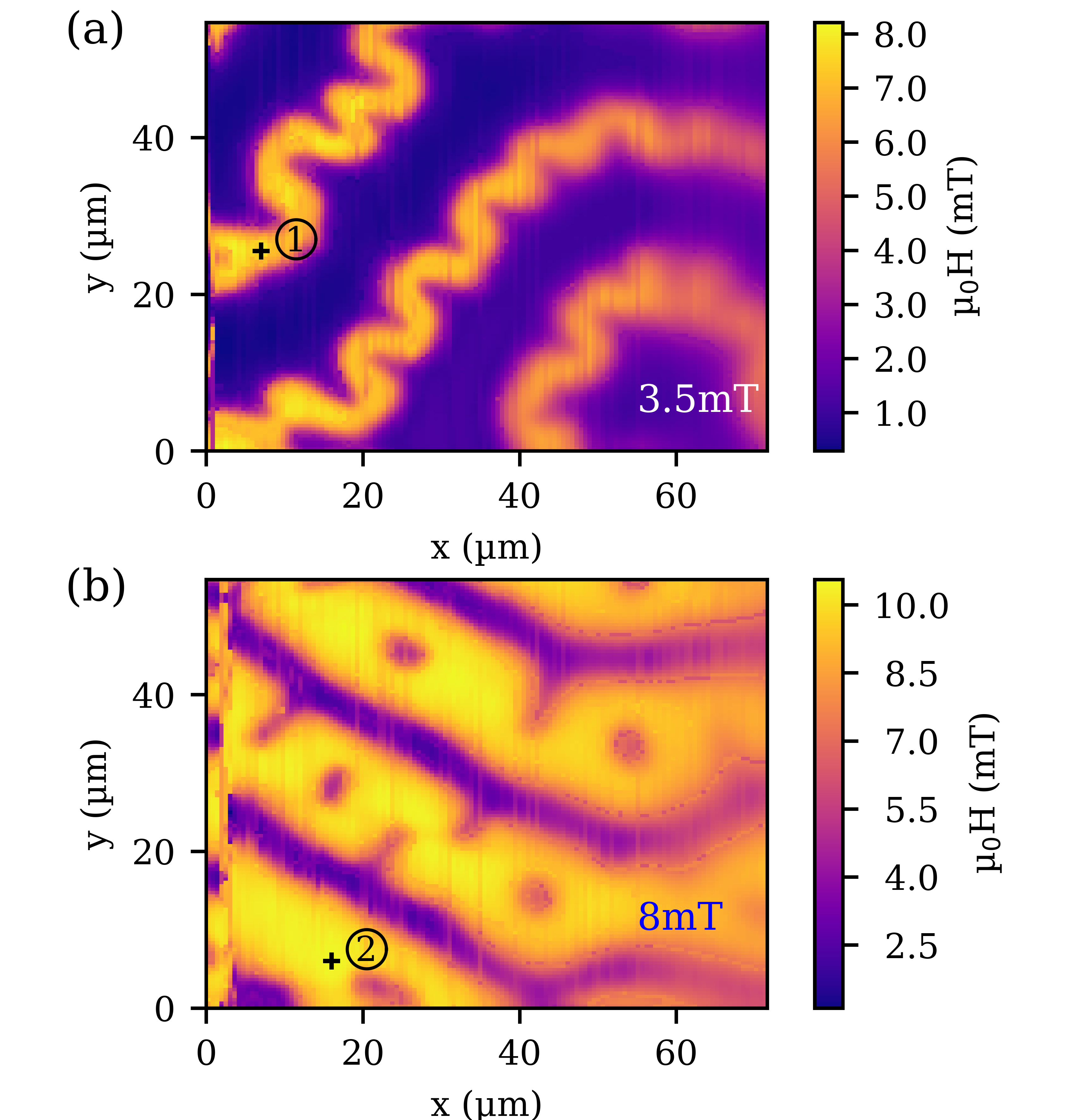}
\caption{(Color online) Scanning SQUID images taken after field cooling under $3.5$~mT (reduced field, $h=H_a/H_c$, of 0.34)  and under $8$~mT (0.61) in (b), at a temperature of $300$~mK, (H$_{c}$=13.1~mT). Points 1 and 2 indicate the maximal measured fields of each scan referred to in Fig.~\ref{fig:PdTe2:BmaxBmin}. The extended domains (open topology) are typical for field cooled type-I superconductors. Dark grey (blue) regions are superconducting and light grey (orange) normal. }
\label{fig:PdTe2:FC}
\end{figure}

\subsection{PdTe$_2$ Field Cooled}

Above we have investigated flux structures in the intermediate state after ZFC. Alternatively, one can reach the intermediate state by field cooling (FC) from the normal phase. Since the obtained magnetic structures sensitively depend on 
 domain wall energy, magnetic-field energy and pinning forces, the intermediate state patterns can be very different~\cite{Huebener,Goren1971,Prozorov2005}. In the FC case, the expulsion of the flux in general results in laminar structures, that connect to the edges of the crystal. Two examples of such open topology structures are presented in Fig.~\ref{fig:PdTe2:FC}, scanned at $0.3$~K in $3.5$~mT and $8$~mT applied fields. The $8$~mT scan shows a mixture of laminar and tubular superconducting structures, while in the $3.5$~mT scan meandering normal state laminae are present.

\section{Discussion}
One of the major results from the present SSM measurements is the direct observation of the intermediate state on a local scale in the field range $(1-N)H_c < H_a < H_c$, with the succession of tubular to laminar structures as the applied field is increased.
According to the theory of the formation of the intermediate state in the bulk of a type-I superconductor the magnetic field in the normal domains should always be equal to $H_c$~\cite{Tinkham}. In the case of PdTe${}_2$ this was demonstrated by $\mu$SR measurements~\cite{Leng2019} probing the field in the bulk of the normal domains in the crystal.

When Landau established the laminar model~\cite{Landau1937} of the intermediate state consisting of alternating superconducting and non-superconducting laminae, he took into account the shape of the laminae close to the sample surface. The magnetic field lines have to bend when they enter via the normal laminae into the sample. Deep inside the normal lamina the flux is compressed and magnetic induction reaches $\mu_{0}H_{c}$. On the other hand the magnetic flux at the outer surface of the normal lamina is less compressed resulting in a magnetic field reduced compared to $H_c$. Based on the model of Landau~\cite{Landau1937} Lifshitz and Sharvin~\cite{Lifshitz1951} calculated this reduction numerically in 1951 and Fortini \textit{et~al.}~\cite{Fortini1972} obtained an analytical expression for this reduction of the magnetic field in the normal domains. 

Though Ref.~\cite{Clem2013} succeeds to identify the energetically most favorable flux configuration, tubular, laminar and again tubular in minimizing the sum of the surface energies between the normal and the superconducting domains and the energy between the sample and the outside space, the authors of Ref.~\cite{Tinkham,Clem2013} mention that they did not consider the deviation from $\mu_{0}H_{c}$ of the magnetic flux density in the normal domains due to the domain wall bending at the sample surface. 

A consequence of the domain wall bending is the reduction of the magnetic induction at the surface of the normal domain. The experimental observations of this are scarce~\cite{Kozhevnikov,Prozorov2005}. Magneto-optical measurements~\cite{Prozorov2005} at 10 $\mu$m above the surface revealed a reduction by a factor 2 of the magnetic induction upon flux entry (tubular phase, closed topology) compared to flux exit (laminar phase, open topology). The authors attribute their finding to the spreading of the flux tubes at the surface. 

Scanning SQUID microscopy allows to quantify the flux entering into the normal laminae. The SQUID detects the perpendicular component of the magnetic induction that threads the SQUID at a given height, in our case 350 nm above the surface. 
Of each image obtained after ZFC at 900~mK we have read the highest value of the magnetic induction after a step wise increase of the applied magnetic field and in the same manner after field cooling to 300~mK.

\begin{figure}[htb!]
	\centering
	\includegraphics[scale=0.95]{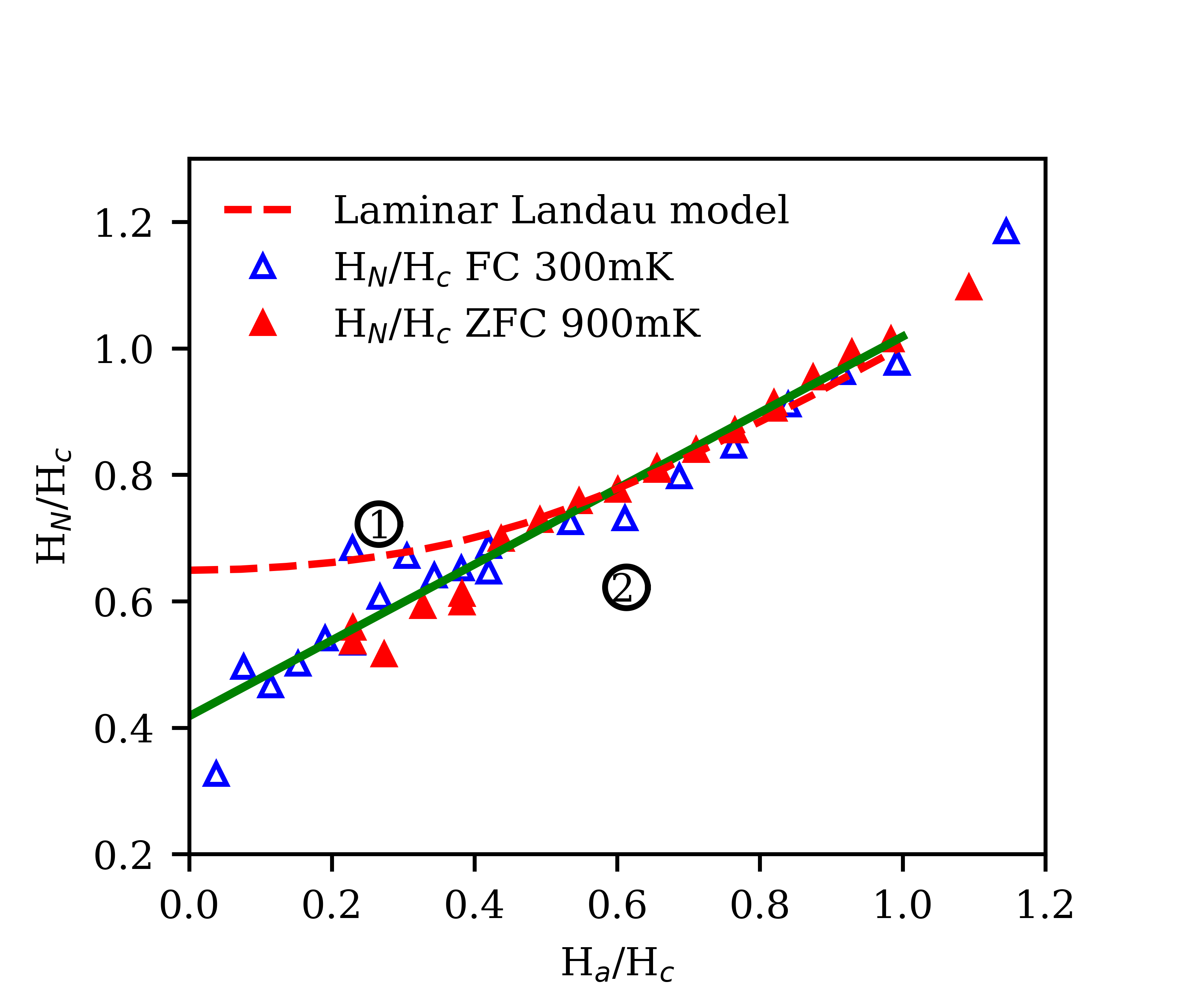}
	\caption{(Color online) The maximum of the magnetic field threading the SQUID loop above the normal regions in the intermediate state divided by the critical field, $H_N$/$H_{c}$ (triangles) as a function of the applied field divided by the critical field, $H_a$/$H_{c}$, for in red/full, ZFC ($900$~mK) and, in blue/empty, FC ($300$~mK) measurements. The green (dark gray) line represents a linear fit. The red dashed line traces the reduction of the maximal field at the surface in case of the Landau laminar model as reported by Fortini \textit{et~al.}~\cite{Fortini1972}. Points 1 and 2 are references to the corresponding points in the images of Fig.~\ref{fig:PdTe2:FC}.} 
	\label{fig:PdTe2:BmaxBmin}
\end{figure}

The result is shown in Fig.~\ref{fig:PdTe2:BmaxBmin} where we trace $H_{N}$, the maximal field of the normal state structure (tubes and laminae), normalized by $H_c$, as a function of the reduced applied field $H_a / H_c$. We observe that $H_N/ H_c$ increases in a linear fashion as function of the applied field. The initial offset corresponds to the highest field measured in the first tubular structure funneling magnetic flux through the sample. For reduced fields $H_N/ H_c > 0.5$ the measured values align with the prediction of Landau for the laminar model as expressed in Fortini~\textit{et~al.} ~\cite{Fortini1972}. We observe at $H_N/ H_c > 0.5$ the formation of laminar structures Fig.~\ref{fig:PdTe2:ZFCimages}(c). At lower fields in the case of zero field cooling tubular structures are present or meandering laminar-like open topology structures for field cooling. As Domain wall bending is expected in either case flux spreading should be observed in tubular and laminar structures.
 
The Landau laminar model, taking into account flux spreading, supposes formation of laminar structures from the onset of the intermediate state. The recent theoretical work of Clem~\textit{et~al.}~\cite{Clem2013} shows that the bending of the normal-superconducting interface between laminae allows for a lower energy state for reduced fields, $H_a / H_c$ \textgreater
0.2, compared to straight interfaces. But nevertheless at lower fields Clem ~\textit{et~al.} show that an array of isolated flux tubes, even with straight interfaces, is energetically favorable compared to laminae with bent interfaces i.e. presenting flux spreading.
Thus a model taking into account bent interfaces for flux tubes would extend the field range for which tubular structures are favored compared to laminar ($H_a / H_c$\textgreater 0.2) structures. 
When tubular structures are energetically favored compared to laminar ones then consequently the magnetic field at the surface will continue to decrease as tubular structures replace laminar ones.
This is supported by our observation (see Fig.~\ref{fig:PdTe2:BmaxBmin}). A complete picture of the magnetic state in type-I superconductors has to expand the models of Goren~\textit{et~al.}~\cite{Goren1971} or Clem~\textit{et~al.}~\cite{Clem2013} by taking into account bent interfaces. For this the work of Fortini~\textit{et~al.}~\cite{Fortini1972} has to be expanded to tubular structures.
 
Our measurement of the magnetic field at the sample surface allows to estimate the degree of spreading based on conservation of flux.
Taking into account conservation of flux in a single tubular normal domain, $\Phi=\mu_0H_c S_{bulk}\approx \mu_0H_{surf} S_{surf}$, where $S_{bulk}$ and $S_{surf}$ are the cross sections of the flux tube in the bulk and at the surface respectively. This implies the ratio $H_{surf}/H_c$ behaves as $S_{bulk}/S_{surf}$. We can consider $H_N=H_{surf}$ as a first approximation and $H_{surf}/H_c \approx 0.5$ at the onset of the intermediate state (see Fig.~\ref{fig:PdTe2:BmaxBmin}), which implies the diameter $d_{surf}$ of a flux tube near the surface being 1.4 times the one in the bulk $d_{bulk}$. By increasing the field this effect becomes smaller and smaller, as the normal state is approached and the energy difference between inside and outside of the sample diminishes.
 
The question how far this spreading effect carries over into the bulk of the superconductor has been calculated ~\cite{Fortini1972} in the framework of the Landau laminar model. The pertinent length scale is $a$, the spacing between the normal domains. Thus for 30\% filling fraction the authors predict a characteristic depth of about 0.20$a$ for the interface bending. The typical spacing of the order of 20~$\mu$m would indicate a bending of the interface between the flux tube and the superconducting phase over at least a depth of 4~$\mu$m.

In the case of a vortex in a type-II superconductor the length scale of spreading \cite{Carneiro2000} of the magnetic field is the penetration depth. The extent and the depth of bending in a type-I superconductor are at least an order of magnitude more important, resulting in a far more spread out field profile. Fitting the flux profile of a flux tube in a type-I superconductor using the model of a vortex for a type-II superconductor does not take into account this difference in spreading. Consequently the model free approach of integrating the flux,Fig.~\ref{fig:PdTe2:ZFCimages}(e) should give a more adequate value of the flux carried in the tube in Fig.~\ref{fig:PdTe2:ZFCimages}(a).

\section{Summary and conclusion}

Using a high resolution scanning $\mu$-SQUID microscope we have investigated the local magnetic flux structure in the intermediate state of the type-I superconductor PdTe$_2$. The data have been taken on a thin single crystal with a demagnetization factor $N=0.788$.
By analyzing the SQUID signal as a function of the applied magnetic field at several fixed temperatures we have determined $(1-N)H_c$ and $H_c$ and obtained the boundaries between the Meissner, intermediate and normal states.
The measured value $H_c = 13.6$~mT is in excellent agreement with the literature~\cite{Carley}.
The success of this approach is the result of a very low resistance to flux penetration and of weak flux pinning in this crystal of PdTe$_2$.
The magnetic images reveal the intermediate state and thus type-I superconductivity. In the intermediate state we also identify a field $H_f$ where tubular, closed topology, flux structures fuse into laminar, open topology, structures. Both type of structures coexist at fields above $H_f$.

We estimated the domain wall width in analyzing the size and the period of the flux structures using the model of Goren and Tinkham~\cite{Goren1971} and the model of Clem~\textit{et al.} ~\cite{Clem2013}. 

Furthermore, we observed magnetic flux spreading at the surface, linked to the bending of the interface between normal and superconducting regions close to the sample surface. Consequently the field measured at the surface of the normal domains is smaller than $H_c$, and shows a linear increase with increasing applied field in agreement with the models of Landau, Sharvin and Fortini~\textit{et al.} established for the laminar state. 

Interface bending is also present in the tubular state thus making it more stable towards the transition to the laminar state. In our case at approx 0.5 $H_a/H_c$ the tubular state transits towards the laminar one.
The smallest magnetic flux structure we observed is carrying $3\Phi_0$ of flux.
Finally, single-quantum vortices and type-II superconducting regions were not detected in our experiment, which excludes type II/1 behavior or a mixture of type-I and type-II behavior. The scanning SQUID data fully support PdTe$_2$ is a clean type-I superconductor with very weak flux pinning.

\section{Acknowledgements}

PGC acknowledges funding from the Innovation Programme under the Marie Sk{\l}odowska-Curie grant agreement No. 754303 and the \textit{Fondation des Nanosciences} (FCSN 2018 02D). The research leading to these results received funding from the European Union’s Horizon 2020 Research and Innovation Program, under Grant Agreement No. 824109, the European Microkelvin Platform (EMP).
We are indebted to Thierry Crozes (NEEL/CNRS) and Arnaud Barbier (IRAM) for device fabrication.


\end{document}